\documentclass[
    ,final            
  ]
  {aipproc}

\layoutstyle{6x9}

\usepackage{amssymb}

\begin{document}

\title{The ULX NGC~1313~X-2 : an optical study revealing an interesting behavior}

\classification{97.80.Jp -- 97.60.Lf -- 97.10.Gz -- 98.20.Bg}
\keywords      {X-ray binaries -- Black Holes -- Accretion disks - OB stars association}

\author{Fabien Gris\'e}{
  address={Observatoire Astronomique de Strasbourg, 11 rue de l'Universit\'e, 67000 Strasbourg, France}
}

\author{Manfred W. Pakull}{
  address={Observatoire Astronomique de Strasbourg, 11 rue de l'Universit\'e, 67000 Strasbourg, France}
}

\author{Roberto Soria}{
  address={Mullard Space Science Laboratory (UCL), Holmbury St Mary, Dorking, Surrey RH5 6NT, UK}
}

\author{Christian Motch}{
  address={Observatoire Astronomique de Strasbourg, 11 rue de l'Universit\'e, 67000 Strasbourg, France}
}

\begin{abstract}
We present a summary of our ongoing efforts to study one of the brightest ultraluminous X-ray source,  NGC~1313~X-2. Despite a large coverage in the X-rays, much of the information we have about the source and its environment comes from optical wavelenghts. Here, we report on the properties of the stellar environment, and the differences in the optical counterpart between our two observing epochs (2003--2004 and 2007--2008). We summarize our ongoing program designed to look for radial velocity variations in the optical spectra and for photometric variability.



\end{abstract}

\maketitle


\section{Introduction}
 Ultraluminous X-ray sources (ULXs) are extragalactic X-ray sources that are not at the nucleus of their galaxy, emitting
well above the Eddington limit of a $10\ \mathrm{M_{\odot}}$ black~hole ($L_\mathrm{X}\sim
10^{39}\ \mathrm{erg\ s^{-1}}$) if we assume that they emit isotropically. An important
question is whether they contain
intermediate mass black holes (e.g. \citep{1999ApJ...519...89C}), whether they are 
beamed \citep{2001ApJ...552L.109K}, or if they are rather 
normal X-ray binaries with super-Eddington emission \citep{2002ApJ...568L..97B}.

NGC~1313~X-2 has been extensively studied in X-rays (e.g. \citep{2006ApJ...650L..75F}), and is also one of the best ULX studied in the optical wavelenghts (\citet{2008A&A...486..151G} and references therein). The recent announcement of a periodic signal in the HST/WFPC2 optical lightcurve of X-2 was interpreted \citep{2009ApJ...690L..39L} as the orbital period.
But despite huge observational efforts, the main parameters of the ULX system remain unknown.

\section{Results}
\subsection{Stellar environment}
Near the ULX, we highlighted \citep{2008A&A...486..151G} two groups of (a few) young stars, spread out 
over $\approx 200$ pc, and hence more similar to an OB association 
(or more likely, two separate associations close to each other) 
rather than to a bound cluster. They clearly stand out in brightness and colors 
over the surrounding old population.
There are no other similar groups of young stars 
in this region of the galaxy, nor are they connected to 
spiral arm features. The reason for this recent, localized episode
of star formation is unclear, but the ULX is clearly associated 
with this young population.


We estimate \citep{2008A&A...486..151G} that the largest association of young stars has an age $\approx 20$ Myr
and a stellar mass $\approx 5 \times 10^3\ \mathrm{M_{\odot}}$. 
The ULX optical counterpart (V $\sim$ 23 mag) appears as one of the brightest stars in the association, 
without any obvious color or brightness anomaly. Using standard stellar evolutionary 
tracks, we constrain its mass to be $\lesssim 12 \mathrm{M_{\odot}}$ ; or even less, if the accretion disk 
is significantly contributing to the source luminosity. We estimate older ages for the stellar association than reported in previous work \citep{2007ApJ...661..165L}, and, correspondingly, 
lower stellar masses (see \citep{2008A&A...486..151G} for more details).

\subsection{Optical counterpart}

Optical spectra of the ULX counterpart taken in 2003--2004
reveal characteristic high-excitation emission lines,
including a broad He{\footnotesize II} $\lambda 4686$ line
(\citep{2006IAUS..230..293P}, Gris\'e et al, in prep.). Such broad line usually comes
from the accretion disk around the black hole, thus confirming
the association with the X-ray source. We detected \citep{2006IAUS..230..293P}
a significant velocity variation of this line
($\Delta v = 380 \pm 30$ km s$^{-1}$) between two spectra
taken at a 3 weeks' interval (Figure 1, left panel).
If this variation reflects the orbital motion of the black hole,
we can rule out the presence of an intermediate-mass black hole
with $M \gtrsim 50\ \mathrm{M_{\odot}}$. Our recent (2007--2008) spectroscopic
campaign on the VLT revealed a decrease of the relative flux
in the He{\footnotesize II} line (Figure 1, right panel).
The line is no longer resolved, and the equivalent width
has decreased from $\approx 10$ \AA~to $\approx 3$ \AA.
This affects the detection of the line in individual
spectra, especially those taken in non-optimal seeing conditions;
thus, it will make it very difficult to constrain the radial
velocity curve.

Another significant finding of our work is the short-term optical
variability of the ULX counterpart during the 2003--2004 campaign \citep{2008A&A...486..151G}, by up to $\approx 0.2$ mag, 
on timescales of hours and days (Figure~\ref{1313x2_photom}, left panel). This is detected 
both in the HST/ACS and in the VLT/FORS1 datasets from 2003--2004, 
and even more evident in the combined dataset.
There is no evidence of periodicity. This suggests that the variability 
is not due to ellipsoidal variations. Instead, it may be caused by varying 
X-ray irradiation of the donor star and (more likely) a stochastically-varying 
contribution from the accretion disk. 
Our new photometric follow-up seems to confirm this interpretation and suggests that the source has entered a different "state", with a mean optical luminosity at a low level 
consistent with the lowest luminosity of the earlier study (Figure~\ref{1313x2_photom}, right panel). The X-ray flux of X-2 about this epoch was
also much lower than the long-term average. The optical flux
is consistent with being constant during the 2007--2008
observations, in contrast with the short-term variations
in 2003--2004. These findings suggest a lower accretion rate
and/or lower radiative efficiency of the ULX, reflected
both in the optical and X-ray flux. In particular, we are
investigating how the decrease in the X-ray flux may reduce
the He{\footnotesize II} $\lambda 4686$ line emission
from the irradiated accretion disk and companion star.

\begin{figure}
\begin{tabular}{cc}
  \includegraphics[width=0.44\columnwidth, bb=28 58 570 600, angle=270]{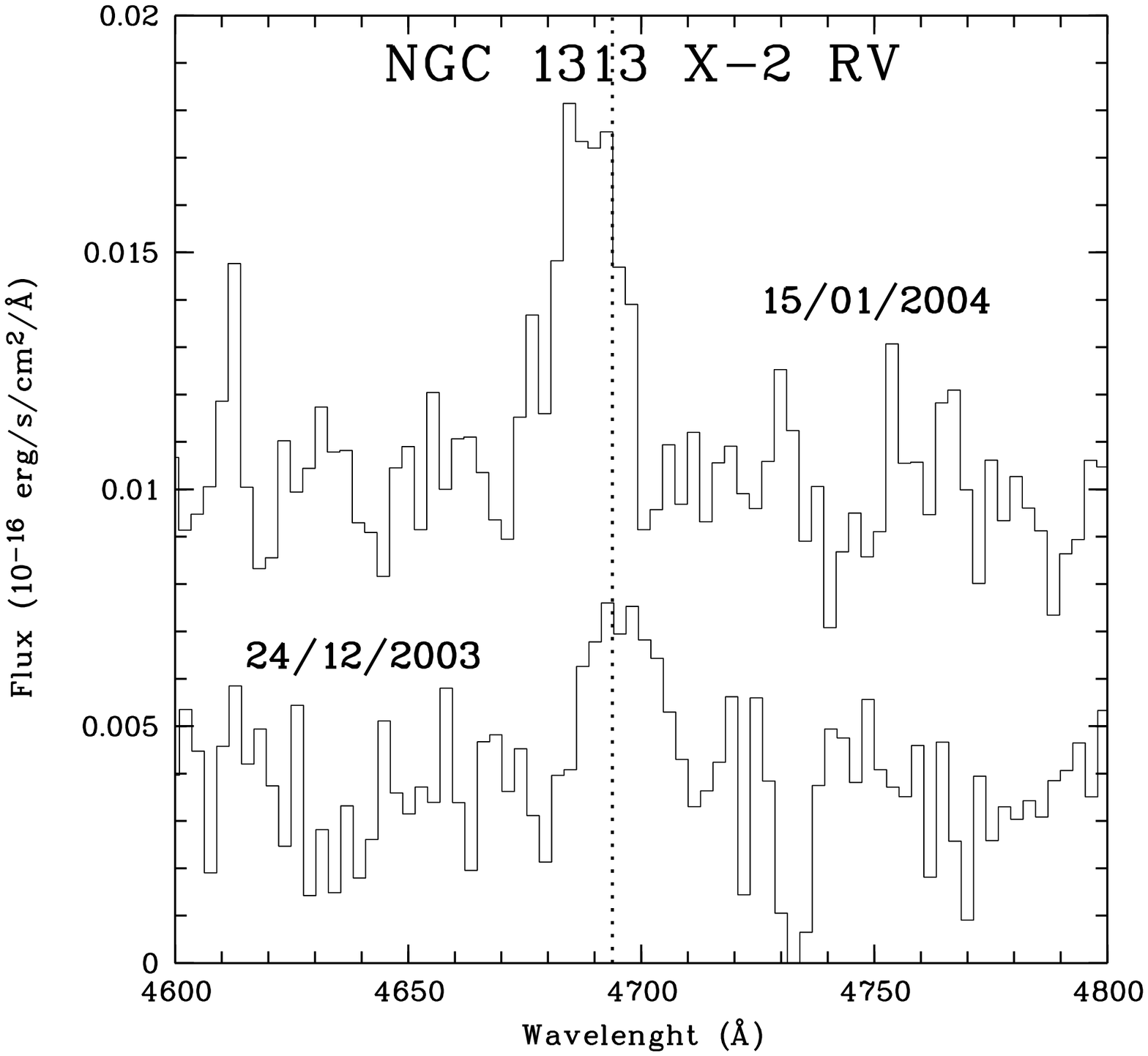}
& \includegraphics[width=0.44\columnwidth, bb=28 28 570 600, angle=270]{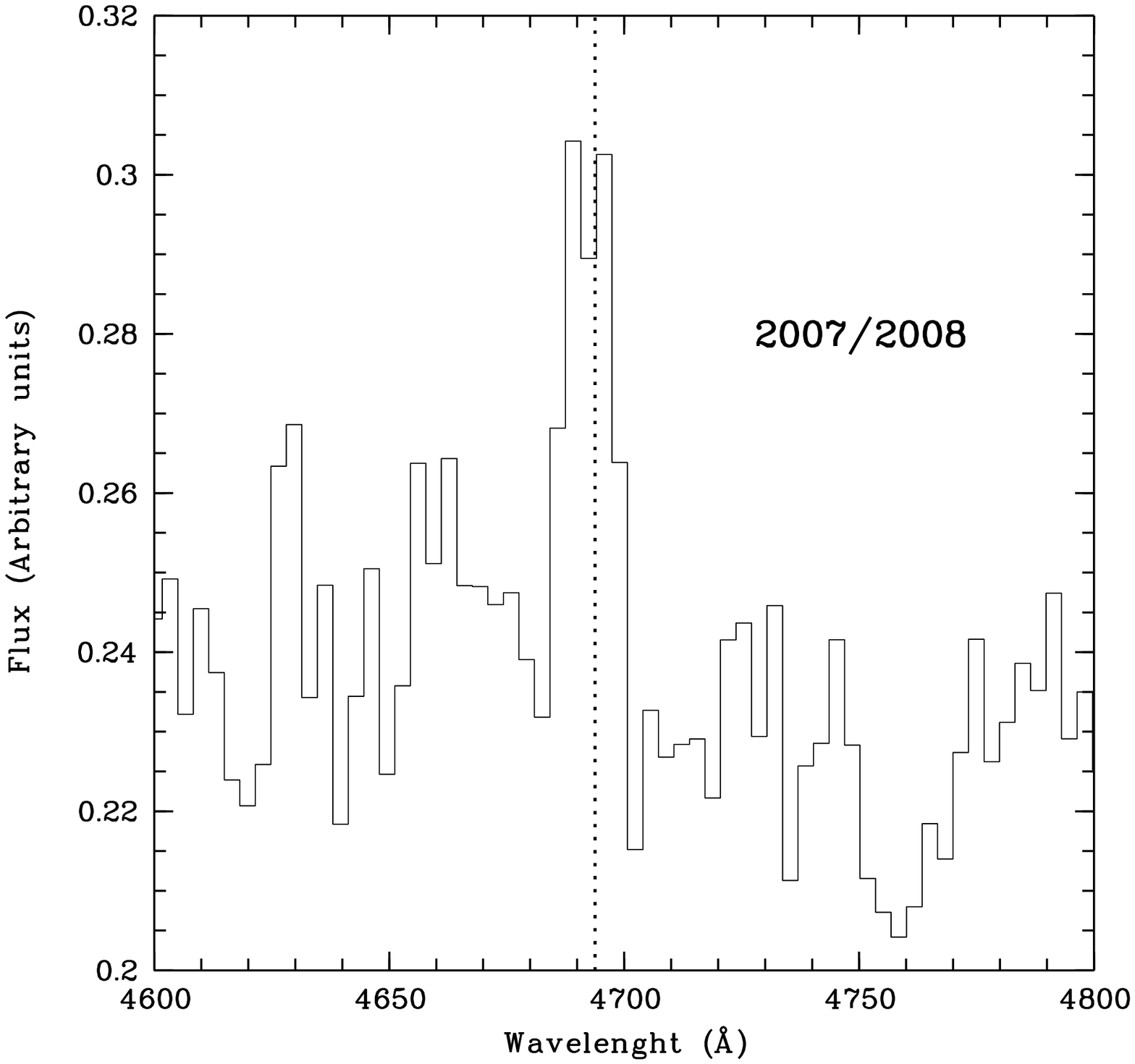}\\
\end{tabular}
  \caption{{\it Left} : radial velocity variation of the He{\footnotesize II} $\lambda 4686$ line of the optical counterpart of X-2.
The two spectra ($2 \times 20$ minutes of exposure time each) show a clear, resolved line (intrinsic FWHM~$\sim$~10~\AA)  and 
display an interesting
variation of $380 \pm 30$ km/s around the systemic velocity of the nebula that surrounds the ULX.
{\it Right} : combined spectrum observed in 2007--2008 ($9 \times 45$ minutes of exposure time) showing an unresolved 
He{\footnotesize II} line (intrinsic FWHM < 5~\AA)  with
a drop in the equivalent width of a factor $\approx 3.5$.}
\label{1313x2_rv}
\end{figure}

\begin{figure}
   \begin{tabular}{cc}
   \includegraphics[width=0.44\columnwidth]{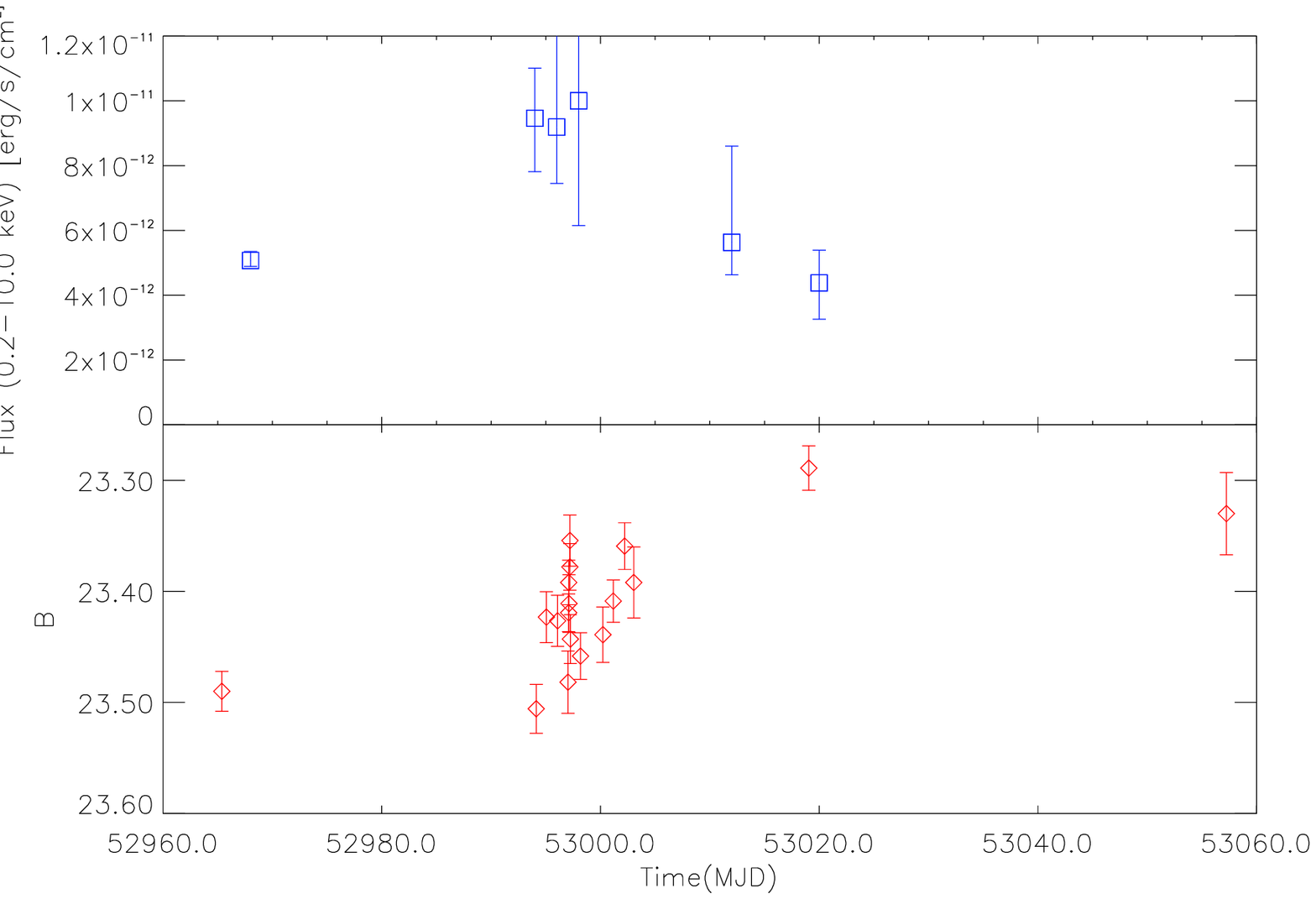}
&  \includegraphics[width=0.44\columnwidth]{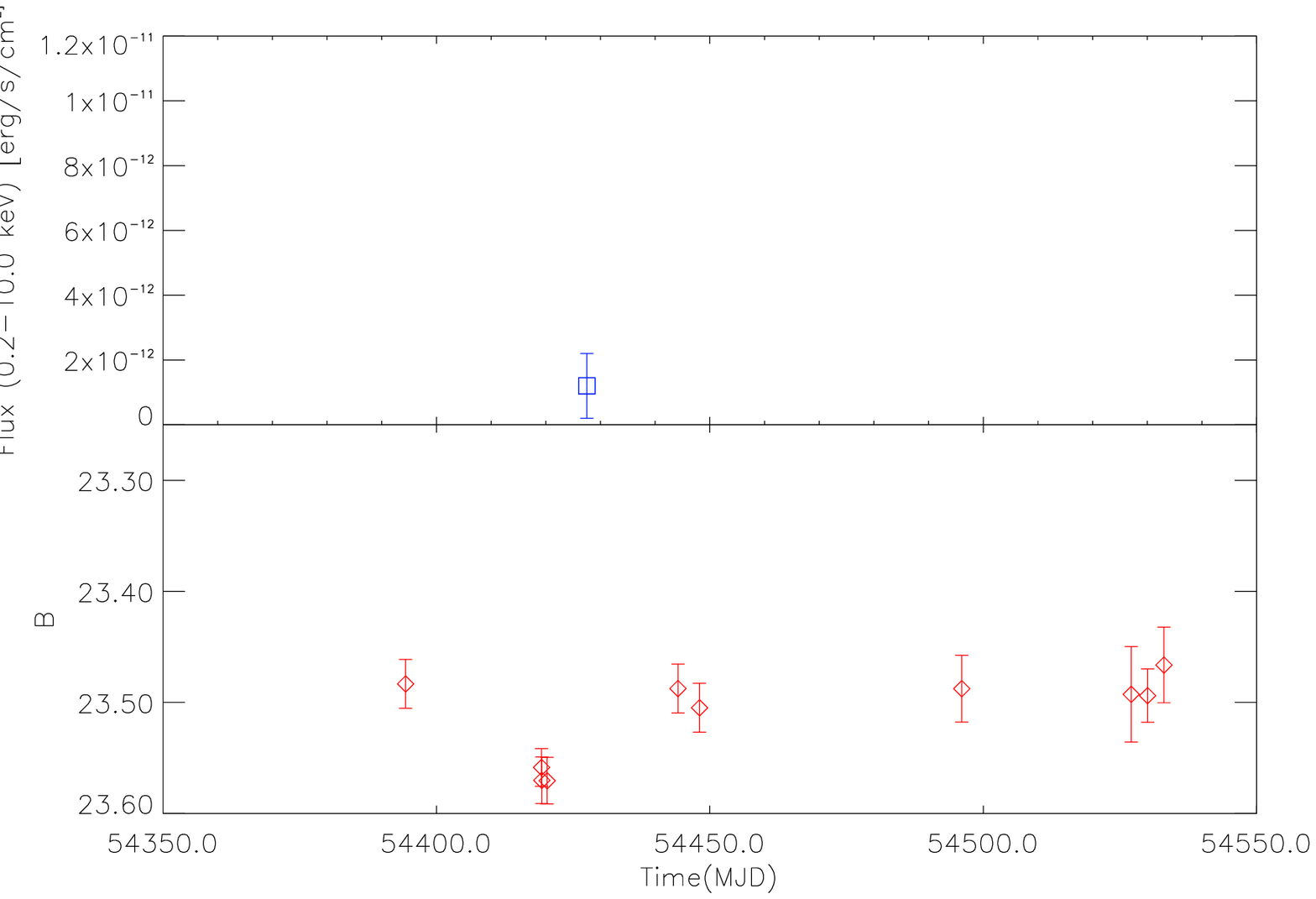}
   \end{tabular}
   \caption{X-ray (top) and $B$-band (bottom) light curve of the ULX counterpart between 
2003 November 22 and 2004 February 22 ({\it left}) and between 2007 October 21 and 2008 March 8 ({\it right}). The optical light curve comes from VLT/FORS1 and HST/ACS data. 
Note that the X-ray flux (unabsorbed) is taken from \citep{2007ApJ...658..999M} for the left plot and from \citep{2008ATel.1530....1P} for the right plot.}
\label{1313x2_photom}
\end{figure}

\bibliographystyle{aipproc}   

\bibliography{f_grise1}

\IfFileExists{\jobname.bbl}{}
 {\typeout{}
  \typeout{******************************************}
  \typeout{** Please run "bibtex \jobname" to optain}
  \typeout{** the bibliography and then re-run LaTeX}
  \typeout{** twice to fix the references!}
  \typeout{******************************************}
  \typeout{}
 }

\end{document}